\documentclass[twocolumn,showpacs,prl]{revtex4-1}

\usepackage{graphicx}
\usepackage{epsfig}
\usepackage{natbib}

\begin{document}

\title{Dissipative soliton excitability induced by spatial inhomogeneities and
drift}
\author{P. Parra-Rivas$^{1,2}$, D. Gomila$^{1}$, M.A. Mat\'ias$^{1}$, and P. Colet$^{1}$}
\affiliation{
$^{1}$ IFISC, Instituto de F\'isica Interdisciplinar y Sistemas Complejos (CSIC-UIB)\\ 
E-07122 Palma de Mallorca, Spain \\
$^{2}$ Applied Physics Research Group (APHY), Vrije Universiteit Brussel, Pleinlaan 2, 1050 Brussels, Belgium}

\begin{abstract}
We show that excitability is generic in systems displaying dissipative solitons
when spatial inhomogeneities and drift are present. Thus, dissipative solitons
in systems which do not have oscillatory states, such as the prototypical
Swift-Hohenberg equation, display oscillations and Type I and II excitability
when adding inhomogeneities and drift to the system. This rich dynamical
behavior arises from the interplay between the pinning to the inhomogeneity and
the pulling of the drift. The scenario presented here provides a general
theoretical understanding of oscillatory regimes of dissipative solitons
reported in semiconductor microresonators. Our results open also the possibility
to observe this phenomenon in a wide variety of physical systems.
\end{abstract}

\maketitle

Spatially extended systems display a large variety of emergent
behaviors \cite{CH}, including coherent structures.
Particularly interesting is the case of dissipative solitons
(DS) \cite{Akhmediev1,*Akhmediev2}, exponentially localized structures
in dissipative systems driven out of equilibrium, as they
can behave like discrete objects in continuous systems. DS
can display a variety of dynamical regimes such as periodic oscillations
\cite{Umbanhowar,Firth1,*Firth2,Vanag}, chaos \cite{Michaelis,Leo12}, or excitability
\cite{Gomila05a,*Gomila05b}. DS emerge from a balance between nonlinearity and spatial
coupling, and driving and dissipation. They are unique once
the system parameters are fixed, and they are different from the
well known conservative solitons that appear as one-parameter
families. 

In optical cavities, DS (also known as cavity solitons) have been proposed as
bits for all-optical memories \cite{Firth02,Barland,Pedaci,Leo10,Odent}, due to
their spatial localization and bistable coexistence with the fundamental
solution. In Ref.~\cite{Gomila05a,*Gomila05b} it was reported that DS may exhibit excitable
behavior. A system is said to be excitable if perturbations below a certain
threshold decay exponentially, while perturbations above this threshold induce a
large response before going back to the resting state. Excitability is found for
parameters close to those where a limit cycle disappears \cite{Izhikevich}.
Excitability mediated by DS is different from the well known dynamics of
excitable media, whose behavior stems from the (local) excitability present in
the system without spatial degrees of freedom. Excitability of DS is an emergent
behavior, arising through the spatial interaction and not present locally.
Moreover, the interaction between different excitable DS can be used to build
all-optical logical gates \cite{Jacobo12}.

In real systems, however, typically solitons are static, so oscillatory or
excitable DS are far from being generic. In this work we present a mechanism
that generically induces dynamical regimes, such as oscillations and excitable
behavior, in which the structure of the DS is preserved. The mechanism relies on
the interplay between spatial inhomogeneities and drift, and therefore can be
implemented under very general conditions. Inhomogeneities, or defects, are
unavoidable in any experimental setup, and drift is also often present in many
optical, fluid and chemical systems, due to misalignments of the mirrors
\cite{Santagiustina,Louvergneaux}, nonlinear crystal birefringence
\cite{walk-off1,*walk-off2}, or parameter gradients \cite{Schapers}, in the first case, and
due to the flow of a fluid in the others \cite{fluids,reactors}. Roughly
speaking, the presence of inhomogeneities and drift introduce two competing
effects. On the one hand, an inhomogeneity pins a DS at a fixed position and, on
the other, the drift tries to pull it out. If the drift overcomes the pinning,
DS solitons are released from the inhomogeneity \cite{Schapers,Caboche1,*Caboche2}. Thus,
depending on the height of the spatial inhomogeneity and the strength of the
drift, we observe three main dynamical regimes: i) stationary (pinned)
solutions, ii) oscillatory regimes, where DS source continuously from the
inhomogeneity, and iii) excitability, where a perturbation may trigger a single
DS that is driven away from the inhomogeneity. This scenario does not
depend on the details of the system nor on the spatial dimensionality. Besides,
it provides a solid theoretical framework to explain the dynamics of DS in
systems with drift, as observed for instance in semiconductor microresonators
\cite{Pedaci,Caboche1,*Caboche2}.

To analyze this scenario we start with the prototypical Swift-Hohenberg
equation (SHE), a generic amplitude equation describing pattern formation in a
large variety of systems \cite{CH,Tlidi,Woods,Burke07}. The SHE is variational,
thus it can not have oscillatory regimes. Adding drift and a spatial
inhomogeneity we have:
\begin{equation}
\displaystyle\frac{\partial u}{\partial t}= -\left(\frac{\partial^{2}}
{\partial x^{2}}+k_{0}^{2}\right)^{2}u+c\frac{\partial u}{\partial x} +
r(x)u+au^{2}-gu^{3}+b(x). 
\label{SHE}
\end{equation}
Here $u(x,t)$ is a real field, $a$ and $g$ real parameters, and $c$ is the
velocity of the drift. The SHE displays DS for $a>\sqrt{27/38}g$
\cite{Woods,Burke07}, so we take $a=1.2$ and $g=1$. We fix $k_0^2=0.5$. We use a
supergaussian gain profile $r(x)=r_0-1+\exp\{-\left[(x-x_{0})/ \epsilon
\right]^{18}\}$ to model a finite system of width $2\epsilon$ so that DS
disappear at the boundaries. We fix $r_0=-0.2$, roughly in the middle of the
subcritical region where DS exist, and, except where otherwise indicated,
$\epsilon=94.0842$. Small changes of the parameter values do not substantially
modify the results. A single spatial inhomogeneity located at the center is
introduced adding a Gaussian profile $b(x)=h\exp\{-\left[(x-x_{0})/\sigma
\right]^{2}\}$ with height $h$ and half-width $\sigma$ \cite{Jacobo08}. This is
motivated by the addressing beams used in optical systems to control DS
\cite{Barland,defects1,*defects2}. We take $\sigma=2.045$, roughly half the width of a DS.
Similar results are found for other values of $\sigma$ provided the width is
small enough to avoid trapping two DS. We have also checked that similar results
are found for a defect in the gain profile $r$. We take $h$ and $c$ as control
parameters \footnote{We consider periodic boundary conditions with a system size
larger than the gain region. Numerical simulations are performed using a
pseudospectral method. Starting from the steady state of a numerical simulation,
bifurcation diagrams are found using a Newton method and continuation techniques
(See Ref. \cite{Gomila05a,*Gomila05b}). Stability is determined from the Jacobian
eigenvalues.}.

\begin{figure}[tbp]
\centering 
\includegraphics{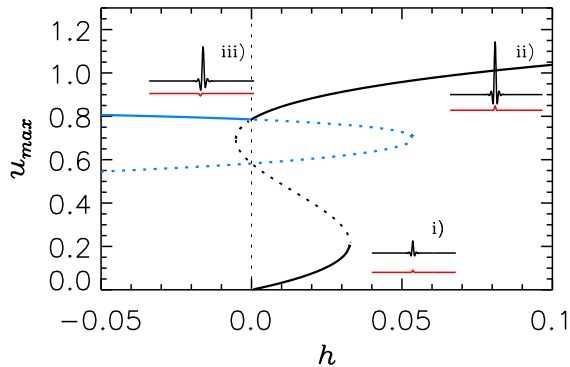}
\caption[9 pt]{(Color online) Bifurcation diagram showing the maximum of the
steady state, $u_{max}$, as a function of $h$ for $c=0$. In black solutions
pinned at their center and in gray (blue) DS pinned at the first tail minima.
Solid (dashed) lines indicate stable (unstable) solutions. The insets show the
profiles of the solution and the spatial inhomogeneity.}
\label{bifdiag0}
\end{figure}

For $c=0$ we observe three different steady state solutions
which are the main attractors of the dynamics (Fig. \ref{bifdiag0}): 
i) the fundamental solution, a low bump corresponding to the
deformation of the homogeneous solution, ii) a high amplitude DS pinned at its
center, and iii) a DS pinned at the first oscillation of its tail. 
Decreasing $h$, at $h=0$, a transcritical bifurcation occurs in which
branch ii) becomes unstable while iii) is stabilized. Physically, at $h=0$ the
defect goes from being a bump to a hole. DS tend to sit at the inhomogeneity
maximum, thus a DS centered at the hole becomes unstable and shifts its
position until the hole coincides with the first minimum of its tail
\footnote{Branch iii) corresponds to pinned DS whose
maximum is at the right of the defect. There is also a degenerated branch with
the maximum at the left. When the drift is turned-on the degeneracy is broken.
We focus in the DS whose maximum is located downstream, since this branch
reconnects to branch ii) when drift is applied.}.
At $h=0$ there is also a crossing of unstable middle-branch
DS. In what follows we focus on $h>0$.

\begin{figure}[tbp]
\centering 
\includegraphics{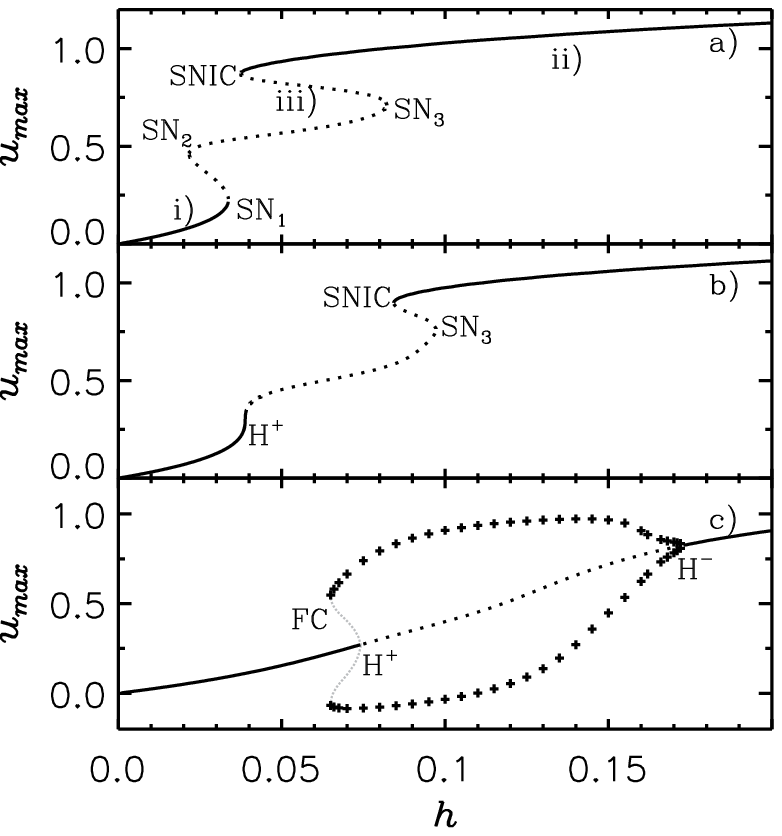}
\caption[9 pt]{The same as in Fig.~\ref{bifdiag0} for a) $c=0.05$, b) $c=0.12$
and c) $c=0.4$ (see main text). In c) the crosses indicate the maximum and
minimum value of the oscillatory DS at a given spatial location. The grey dotted
line is a sketch of the unstable cycle for illustrative purposes.}
\label{bifdiagc}
\end{figure}

\begin{figure}[tbp]
\includegraphics{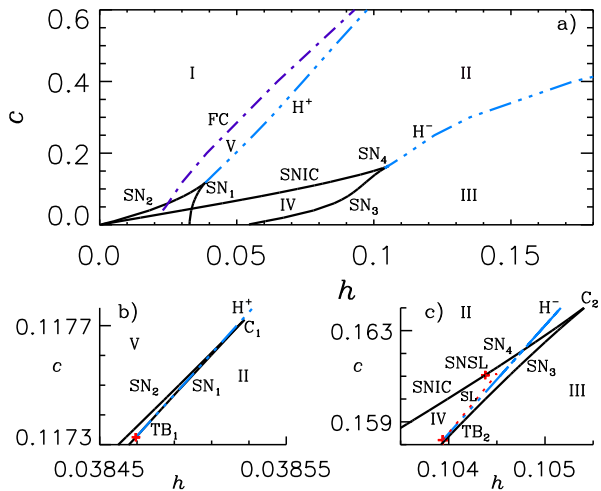}
\caption[10 pt]{a) (Color online) Two-parameter ($c$ vs $h$) phase diagram  of
the system (see text). Panels b) and c) show a zoom close to the cusps and
Takens-Bogdanov points.} \label{F1}
\end{figure}

\begin{figure}[tbp]
\includegraphics{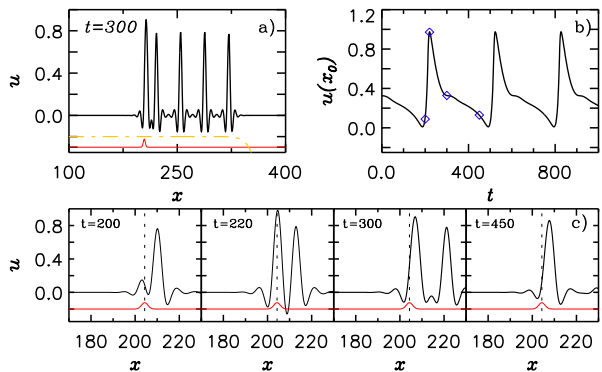}
\caption[10 pt]{a) (Color online) Snapshot of a train of solitons in region II
of Fig.~\ref{F1} for $c=0.11 $ and $h=0.076$. The solid grey (red) and the
dot-dashed lines show $b(x)$ and $r(x)$ respectively. Here $\epsilon=163.625$.
b) Time evolution of the field $u$ at $x_0$. c) Field profile at the times shown
with symbols in b) showing how a new DS develop and detach.}
\label{oscillations}
\end{figure}

This scenario changes when drift is introduced, $c\neq 0$: parity symmetry,
$x \leftrightarrow -x$, is broken making the transcritical bifurcations
imperfect. This implies a rearrangement of branches, that for low
values of $c$ leads to the snake-like branch shown in
Fig.~\ref{bifdiagc}a). Saddle-node bifurcations SN$_1$ and SN$_3$ where
already present for $c=0$. A Saddle-Node on the Invariant Circle (SNIC)
reconnects branches ii) and iii), while the saddle-node SN$_2$ arises from the
middle branches transcritical bifurcation.
Increasing $c$ the branch stretches (cf. Fig.~\ref{bifdiagc}b) and SN$_1$
coalesces with SN$_2$ at the cusp bifurcation C$_1$ as shown in Fig.~\ref{F1}
which displays the general scenario in parameter space. 
Close to C$_1$ there is a Takens-Bogdanov point TB$_1$ that unfolds a Hopf
bifurcation line H$^+$  \cite{Lendert} (Fig.~\ref{F1}b). As $c$ keeps
increasing, the SNIC line encounters a Saddle-Node Separatrix Loop (SNSL)
codimension 2 bifurcation from which a saddle-node SN$_4$ and a saddle-loop SL
bifurcation lines unfold (Fig.~\ref{F1}c). SN$_4$ soon coalesces with SN$_3$ at
the cusp C$_2$ while SL ends at a Takens-Bogdanov point, TB$_2$, which also
unfolds a Hopf line H$^-$ tangent to the SL line. Finally for larger values of
$c$ there is a single monotonous branch of steady state solutions
(Fig.~\ref{bifdiagc}c).
The Hopf line H$^+$ is subcritical and is accompanied by a fold of cycles (FC)
from which a stable cycle emerges. For $h$ large enough the defect is above the
threshold to switch on a DS. Thus this limit cycle corresponds to the periodic
creation of DS at the inhomogeneity that are then drifted away
(Fig.~\ref{oscillations}c), generating a train of solitons that, in our case,
disappear at the boundary of the system (Fig.~\ref{oscillations}a). Such regime
was observed experimentally in semiconductor microresonators \cite{Caboche1,*Caboche2},
where it was called a ``soliton tap''. This cycle is stable all the way to the
supercritical Hopf H$^{-}$, where $h$ becomes large enough to prevent the DS
being advected by the drift, and the cycle ends.

In parameter space (Fig.~\ref{F1}) the fundamental solution is stable in regions
I and V, pinned DS are stable in III and IV and stable trains of DS exist in II
and V. The scenario shown in Figs.~\ref{F1}b and \ref{F1}c, much richer than in
\cite{Gomila05a,*Gomila05b}, is characteristic of systems displaying relaxation oscillations
described, for instance, by the Van der Pol equation (see Fig.~2.1.2 in
\cite{Guckenheimer}). These systems  are characterized by two different time
scales, a slow and a fast one. Here the fast time scale corresponds to the
switching time of a DS on the inhomogeneity, while the slow one is the time the
drift takes to detach a DS once it is formed \cite{Caboche1,*Caboche2}. The switch-on time
is basically independent on the drift, while the detach time has a strong
dependence on it. These two time scales can be clearly identified in a time
trace of the field at $x_0$ (Fig.~\ref{oscillations}b). 
Excitability can be found in regions I, III and IV when applying a
perturbation for a short time that changes either $c$ or $h$. If the system is
brought monetarily into the oscillatory region II then an excitable excursion
consisting on the exploration of a loop of the cycle before returning to the
original steady state is triggered. An alternative way to trigger an excitable
excursion is perturbing the state of the system rather than a parameter
\cite{Izhikevich}. Transient parameter changes are usually easier to implement
experimentally, thus in what follows we consider this kind of perturbations.

\begin{figure}[tbp]
\centering 
\includegraphics{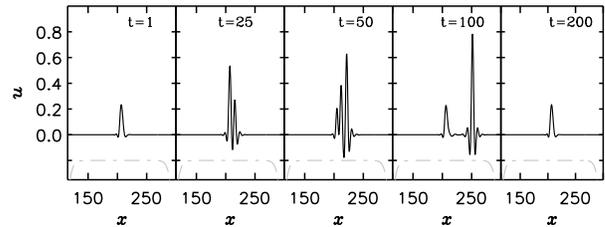}
\caption[9 pt]{Excitable excursion of the fundamental solution in
region I, close to FC ($c=0.6$, $h=0.085$), triggered varying $h$ an amount
$\Delta h = 0.0325$ during $\Delta t = 20$ time units.
The dot-dashed line shows the gain profile $r(x)$.}
\label{exc1}
\end{figure}

\begin{figure}[tbp]
\centering 
\includegraphics{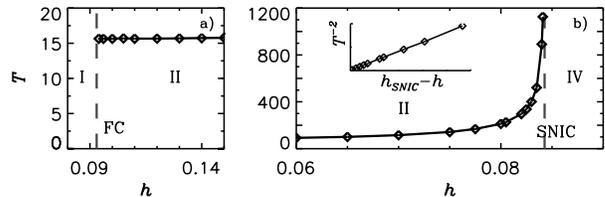}
\caption[9 pt]{Oscillation period $T$ in region II as a function of $h$ close to a)
the FC ($c=0.6$) and b) the SNIC ($c=0.12$). The inset in b) shows the scaling
of the divergence of the period.} \label{periodI}
\end{figure}

In region I a superthreshold perturbation
of the fundamental solution grows to generate a DS, that is
then advected away, setting the defect back to the resting state (Fig.~\ref{exc1}). As the transition from the stationary state to the oscillatory one is mediated by a
subcritical Hopf bifurcation, the excitability is of Type
II \cite{Izhikevich}. A clear signature of this type of excitability is that
the period of the oscillations remains practically constant as one approaches
the threshold from the oscillatory side, namely as one approaches FC coming from
region II, as shown in Fig.~\ref{periodI}a. As typical for Type II excitability
the threshold is rather a 
quasi-threshold. If one applies a perturbation that crosses FC but not
H$^+$, the system will be stacked in region V (bistable). The closer the
parameters are to FC, and the closer is FC to H$^+$, the smaller is the
perturbation required to trigger an excitable excursion 
\footnote{The system is excitable even for $h=0$, although in this
case very large perturbations are required to trigger an excitable excursion.}.

\begin{figure}[tbp]
\centering 
\includegraphics{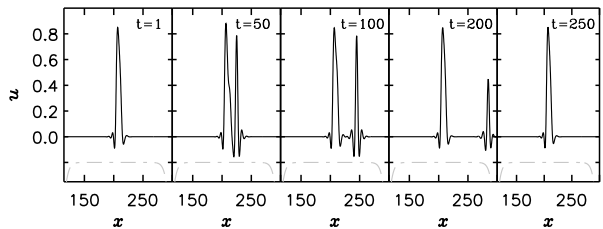}
\caption[9 pt]{The same as in Fig.\ref{exc1} but close to H$^-$, in region III.
Here $c=0.4$, $h=0.18$, and $\Delta h= -0.045$.}
\label{exc2}
\end{figure}

Different dynamics is found in region III, where $h$ is large enough
that, beyond switching on a DS if there is none, it also pins it once it is
formed. In this case DS undergo a supercritical Hopf bifurcation when crossing
H$^-$. Close to threshold, DS exhibit small oscillations, but decreasing $h$ or
increasing $c$ just a little further DS start to source from the inhomogeneity,
forming a sort of {\it canard} in phase space \cite{refCanards}, leading to the
train of DS found in region II. This is again a mechanism leading to Type II
excitability as shown in Fig. \ref{exc2}. At difference with the previous case,
here the resting state is a high amplitude DS (see first panel), and the
excitable excursion consists of the DS leaving the inhomogeneity and a new one
being formed.

There is a third mechanism, associated to the SNIC line separating regions II
and IV. A detailed analysis of DS excitability mediated by a SNIC can be found
in Ref. \cite{Jacobo08}. For parameters in region IV a supra-treshold
perturbation that brings the system into region II triggers the unpinning of a
DS leading to an excitable excursion. The initial and the final state is a
pinned DS and the observed behavior is very similar to the one shown in
Fig.~\ref{exc2}. In this case excitability is of Type I, whose signature is a
divergence of the period. As shown in Fig. \ref{periodI}b) the period diverges
with a power law of exponent $-1/2$ approaching the SNIC line from region II. A
divergence in the period has been reported in semiconductor microresonators
\cite{Caboche1,*Caboche2}, induced also through spatial inhomogeneities and drift.
Additionally, the SL line in Fig.~\ref{F1}c) leads to another Type I
excitable region (discussed at length in \cite{Gomila05a,*Gomila05b}), but at difference
with the SNIC bifurcation line, that extends for a wide parameter range, the SL
occurs in a very narrow region (Fig. \ref{F1}c), and therefore it may be more
difficult to find experimentally.

In summary, we have shown that the competition between the pinning of DS to a
spatial inhomogeneity and the pulling generated by a drift leads to a complex
behavior leading to oscillations and to Type I and II excitability through
several mechanisms. To clearly illustrate this 
we make use of the SHE, for which there is a Lyapunov potential
\cite{CH}, thus the system always evolves to steady states that minimize the
potential. When drift and boundary conditions are added, the SHE no longer
has a Lyapunov potential \cite{Chomaz}, thus complex
dynamical behavior can arise. In order to confirm the generality of our
results, we have checked that the scenario is qualitatively the same in a
completely different model displaying DS \cite{Gomila07}, namely the
Lugiato-Lefever equation \cite{LugiatoLefever}, which has been recently used to
describe DS observed experimentally \cite{Leo10,Odent}. Our results do not
depend on the microscopic details of the system, or the number of spatial
dimensions, or the nature of the spatial inhomogeneity, but on general
emergent properties of DS, providing a theoretical framework to explain the
dynamics of DS in presence of spatial inhomogeneities and drift in a very broad
class of extended systems. The theory presented here gives an explanation to
experimental observations \cite{Caboche1,*Caboche2} and opens up the possibility to observe
this phenomenon in a variety of other experimental setups.

Financial support from the MINECO (Spain) and FEDER under grants
FIS2007-60327 (FISICOS) and TEC2009-14101 (DeCoDicA), FIS2012-30634
(INTENSE@COSYP), and TEC2012-36335 (TRIPHOP), from 
CSIC (Spain) under grants 200450E494 (Grid-CSIC) and PIE-201050I016, and
from Comunitat Aut\`onoma de les Illes Balears is acknowledged.

\bibliography{SHdefectdriftsub}

\end{document}